\documentclass[journal]{IEEEtran}
\usepackage{amsmath, amssymb, amsfonts, amssymb, amsthm}
\usepackage{bbm}
\usepackage{graphicx} \graphicspath{ {figure/} }
\usepackage[noabbrev]{cleveref}
\usepackage{scalefnt}
\usepackage{indentfirst}
\usepackage{cite}
\usepackage{bm}
\usepackage{enumerate}
\usepackage{enumitem}
\usepackage[caption=false,subrefformat=parens,labelformat=parens]{subfig}
\usepackage{scalerel}  
\usepackage{algpseudocode}
\usepackage{etoolbox}
\usepackage{multirow}
\usepackage{caption}
\usepackage[table,xcdraw]{xcolor}
\usepackage{algorithm}
\usepackage{algcompatible}

\renewcommand{\algorithmiccomment}[1]{\hfill ~#1}
\algnewcommand\INPUT{\item[\textbf{Input:}]}
\algnewcommand\INITIAL{\item[\textbf{Initialization:}]}
\algnewcommand\OUTPUT{\item[\textbf{Output:}]}
\algnewcommand\RETURN{\item[\textbf{Return:}]}
\algnewcommand\ITER{\item[\textbf{Iteration:}]}
\algrenewcommand\algorithmiccomment[2][\small]{{#1\hfill\ #2}}

\theoremstyle{plain}

\makeatletter
\newcommand*{\algrule}[1][\algorithmicindent]{%
  \makebox[#1][l]{%
    \hspace*{.2em}
    \vrule height .75\baselineskip depth .25\baselineskip
  }
}

\newcount\ALG@printindent@tempcnta
\def\ALG@printindent{%
    \ifnum \theALG@nested>0
    \ifx\ALG@text\ALG@x@notext
    \else
    \unskip
    \ALG@printindent@tempcnta=1
    \loop
    \algrule[\csname ALG@ind@\the\ALG@printindent@tempcnta\endcsname]%
    \advance \ALG@printindent@tempcnta 1
    \ifnum \ALG@printindent@tempcnta<\numexpr\theALG@nested+1\relax
    \repeat
    \fi
    \fi
}
\patchcmd{\ALG@doentity}{\noindent\hskip\ALG@tlm}{\ALG@printindent}{}{\errmessage{failed to patch}}
\patchcmd{\ALG@doentity}{\item[]\nointerlineskip}{}{}{} 
\makeatother

\renewcommand{\algorithmiccomment}[2][.5\linewidth]{\leavevmode\hfill\makebox[#1][l]{//~#2}}

\theoremstyle{definition}

\begin{document}
\title{VOMTC: Vision Objects for Millimeter and Terahertz Communications}
\author{Sunwoo Kim\IEEEauthorrefmark{1}, Yongjun Ahn\IEEEauthorrefmark{1}, Daeyoung Park\IEEEauthorrefmark{2}, and Byonghyo Shim\IEEEauthorrefmark{1}\\
\IEEEauthorblockA{\IEEEauthorrefmark{1}Institute of New Media and Communications and Department of Electrical and Computer Engineering, Seoul National University, Korea}\\
\IEEEauthorblockA{\IEEEauthorrefmark{2}Department of Electrical and Computer Engineering, Inha University, Korea}\\
Email: \IEEEauthorrefmark{1}\{swkim, yjahn\}@islab.snu.ac.kr, 
\IEEEauthorrefmark{2}dpark@inha.ac.kr, bshim@snu.ac.kr

\thanks{A part of this paper has been presented at IEEE International Conference on Communications (ICC), Denver, USA, 2024 [1]. This work was supported in part by Institute of Information \& communications Technology Planning \& Evaluation (IITP) grant funded by the Korea government(MSIT) (No. 2021-0-00972, Development of Intelligent Wireless Access Technologies), in part by the National Research Foundation of Korea (NRF) Grant through the Ministry of Science and ICT (MSIT), Korea Government (under Grant 2022R1A5A1027646), and in part by Institute of Information \& communications Technology Planning \& Evaluation (IITP) grant funded by the Korea government(MSIT) (No.RS-2022-00155915, Artificial Intelligence Convergence Innovation Human Resources Development (Inha University)).}}
\maketitle
\begin{abstract}
Recent advances in sensing and computer vision (CV) technologies have opened the door for the application of deep learning (DL)-based CV technologies in the realm of 6G wireless communications.
For the successful application of this emerging technology, it is crucial to have a qualified vision dataset tailored for wireless applications (e.g.,  RGB images containing wireless devices such as laptops and cell phones).
An aim of this paper is to propose a large-scale vision dataset referred to as Vision Objects for Millimeter and Terahertz Communications (VOMTC).
The VOMTC dataset consists of 20,232 pairs of RGB and depth images obtained from a camera attached to the base station (BS), with each pair labeled with three representative object categories (person, cell phone, and laptop) and bounding boxes of the objects.
Through experimental studies of the VOMTC datasets, we show that the beamforming technique exploiting the VOMTC-trained object detector outperforms conventional beamforming techniques.
\end{abstract}
\vspace{-0.3cm} 
\section{Introduction}\label{sec:intro}

In recent years, deep learning (DL) has achieved remarkable success in the computer vision (CV) field, as it can extract useful features from images and videos and leverage them to solve various tasks such as semantic segmentation, image classification, object detection, to name just a few [1], [2].
Motivated by the success of DL-based CV, various approaches have been proposed in recent years to exploit CV in wireless tasks. Notable ones include beamforming, blockage detection, moving object tracking, and localization of mobiles [3], [4].
In contrast to the information acquired from radio frequency (RF) signaling, CV technique can provide the spatial information (e.g., distance, angles, and velocity) and semantic information (e.g., object, crowdness, urgent/dangerous situation).
For example, the DL-based object detector returns the target vehicle's location which can be used to generate a beam heading toward the receiver at vehicle.
Also, the CV-aided crowd estimator returns the user density of a designated area using which the base station (BS) can make the user association or cell switching decision.

While the CV-based wireless systems can offer various benefits over the conventional systems, since it is data-driven in nature, the performance depends heavily on the training dataset.
Recently, there have been some studies proposing datasets targetted for the DL-based wireless tasks [5]-[8].
In [6], light detection and ranging (LiDAR) data collected for the line-of-sight detection and the millimeter wave (mmWave) beam selection has been proposed.
In [7], depth images of the indoor communication environment collected using RGB-d camera has been proposed.
Also, in [8], pairs of wireless and visual data acquired for the received power prediction has been proposed.
While these datasets are useful for the target scenario (e.g., the urban canyon scenario containing vehicles of distinct sizes at daytime [6]), they might not fully characterize the real wireless environments, such as the variation in ambient, diffuse, and specular light over time.
Furthermore, since these datasets mainly contain large-scale objects (e.g., cars, trucks, and buses), one cannot capture the small-scale wireless objects (e.g., the cell phone or laptop).

\begin{figure}[t]
	\centering
	\includegraphics[width=1.0\columnwidth, height = 6.0cm]{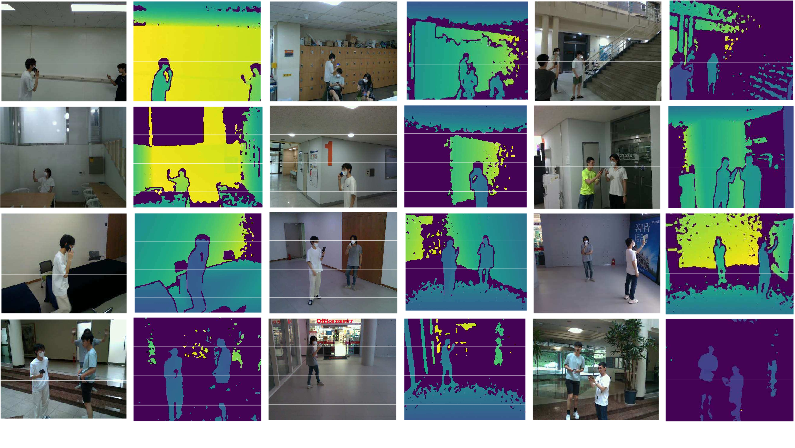}
	\vspace{-0.7cm}
	\caption{Collected RGB-depth image pair samples of VOMTC. Sample images are taken from 2,083 distinct environments.}
 \vspace{-0.5cm}
\label{fig:vobemts_dataset}
\end{figure}

For the successful application of the DL-based CV in wireless communication tasks, one needs to collect dataset from various wireless environments (e.g., office, subway station, hallway, and stadium).
In fact, popularly used CV dataset contains a large number of images acquired from the various environments.
For example, well-known vision datasets such as PLACES [9] and LSUN [10] consist of images collected from 205 and 10 scene categories, respectively.
Since many objects of such image datasets (e.g., chair, horse, dog, and bird) are unrelated to wireless tasks and further these datasets do not contain wireless objects such as the cell phone or laptop, conventional datasets might not be used for training wireless tasks.

Our main contribution in this work is to propose a vision dataset for the CV-based wireless applications.
Our dataset, henceforth referred to as the \textit{Vision Objects for Millimeter and Terahertz Communications} (VOMTC), consists of 20,232 pairs of RGB and depth images labeled with three representative object categories (person, cell phone, and laptop).
In our measurement campaign, we collected samples from 2,083 distinct environments including classroom, cafeteria, elevator, hallway, and stairs and at various times of the day (e.g., morning, afternoon, and night).
In Fig. 1, we illustrate the snapshots of VOMTC (see https://github.com/islab-github/VOMTC).
In order to help the wireless researchers pick the inputs and outputs fitting to their specific needs and applications, we design the dataset selection code\footnote{To download the selection code \texttt{VOMTCdatasetSelection.py}, check out https://github.com/islab-github/VOMTC.}.
By adjusting three key parameters: \textit{1) active classes}, \textit{2) maximum number of people}, and \textit{3) maximum distance to the farthest object}, user can easily extract the desired dataset.
For example, if the target task is to detect laptops in an image, one can set the \textit{active classes} to `laptop' to obtain a dataset consisting of laptops exclusively.


To demonstrate the effectiveness of our dataset, we design a CV-aided beam management technique for mmWave and THz communication systems called VOMTC-aided beam management (VBM). 
Using the RGB image captured at the BS, VBM identifies the cell phone's location which can be used to generate a directional beam heading toward the cell phone.
In our experimental studies, instead of relying on the off-the-shelf object detector  (e.g., Faster-RCNN, YOLO, and EfficientDet [11]-[13]) designed to identify various objects (e.g., cows, elephants, tennis rackets, and kites), we exploit the VOMTC-trained object detector to focus on identifying cell phones.
To come up with such detector, we use EfficientDet as a baseline and then perform the fine-tuning (FT) using the VOMTC dataset.
Due to the use of special object detector tailored for wireless device identification, we can enhance the chance of transmitting directional beams to the phone location.
From the simulation results, we demonstrate that VBM achieves more than 36\% improvement in the data rate over the 5G beam management (5G-BM) technique.

Recently, various attempts have been made to support 
CV-aided beamforming with dataset generation.
In [14], a synthetic multi-modal dataset consisting of image and channel pairs generated from the autonomous driving and ray-tracing simulation software has been proposed to support beamforming in the V2X commmunication scenario.
In [15], a synthetic dataset containing triplets of 3D point clouds, distributions of non-line-of-sight (NLoS) users within the BS coverage range, and target codeword angles has been proposed to support a  beamforming codebook design.
In [16], a dataset of 1,000 images collected for training the off-the-shelf object detector to detect horn antennas in IRS-based systems has been proposed.

The main contributions of this paper, when compared to the conventional CV-aided beamforming approaches [14]-[16], are summarized as follows:
\begin{itemize}
    \item Unlike the datasets in [14]-[16] containing many large-scale objects such as cars, trucks,
    and persons, VOMTC has labels on the small-scale wireless objects (e.g., the cell phone or
    laptop). Since these objects are the real targets for wireless communications, we can perform
    various wireless tasks such as beamforming, random access, and semantic compression.
    For example, VOMTC can be used to design a crowd estimator specialized for the accurate
    density estimation of wireless objects. 
    It can also be used to extract an essential information
    in a sensing image (i.e., class, distance, angles, and number of wireless objects).

    \item  VOMTC is dearly distinct from conventional datasets in [14], [15] in that it offers images
    obtained directly from real wireless environments. We note that the synthetic datasets
    in [14]. [15] might not fully characterize the real wireless environments so that they might
    result in a degraded performance in the test (i.e., beamforming) phase. While the dataset
    developed in [16] has real images, it contains images of horn antennas exclusively so that
    it cannot be used to identify the small-scale wireless objects (e.g., cell phones and laptops).

    \item   While approaches in [14]-[16] rely on the codebook-based beam transmission, we do not exploit the codebook since we directly identify the exact beam direction from the sensing information.
    In fact, from the captured image, a location of the cell phone can be identified via the VOMTC-trained object detector, which can be used to generate a beam heading toward
    the derived location. 
    Whereas, approaches in [14]-[16] select the beam codeword from the predefined codebook so that the mismatch between the pre-defined beam direction and the real one is unavoidable.

\end{itemize}

\begin{figure}
\centering
\subfloat[]{\includegraphics[angle=0,origin=c,width=0.4\columnwidth, height=5.0cm]{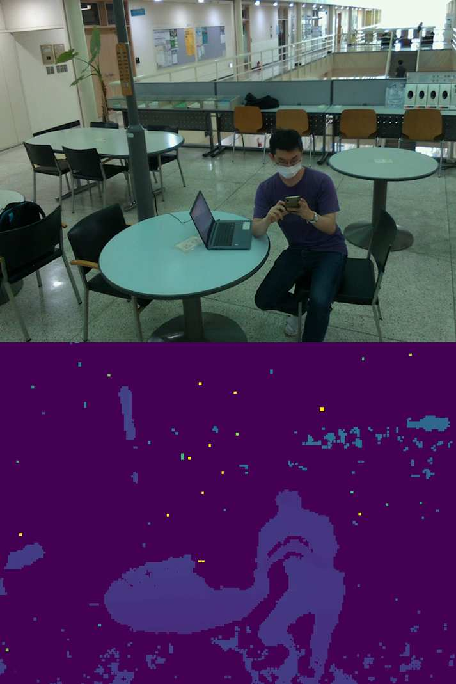}}
	\subfloat[]{\includegraphics[angle=270, origin=c, width=0.6\columnwidth, height=5.0cm]{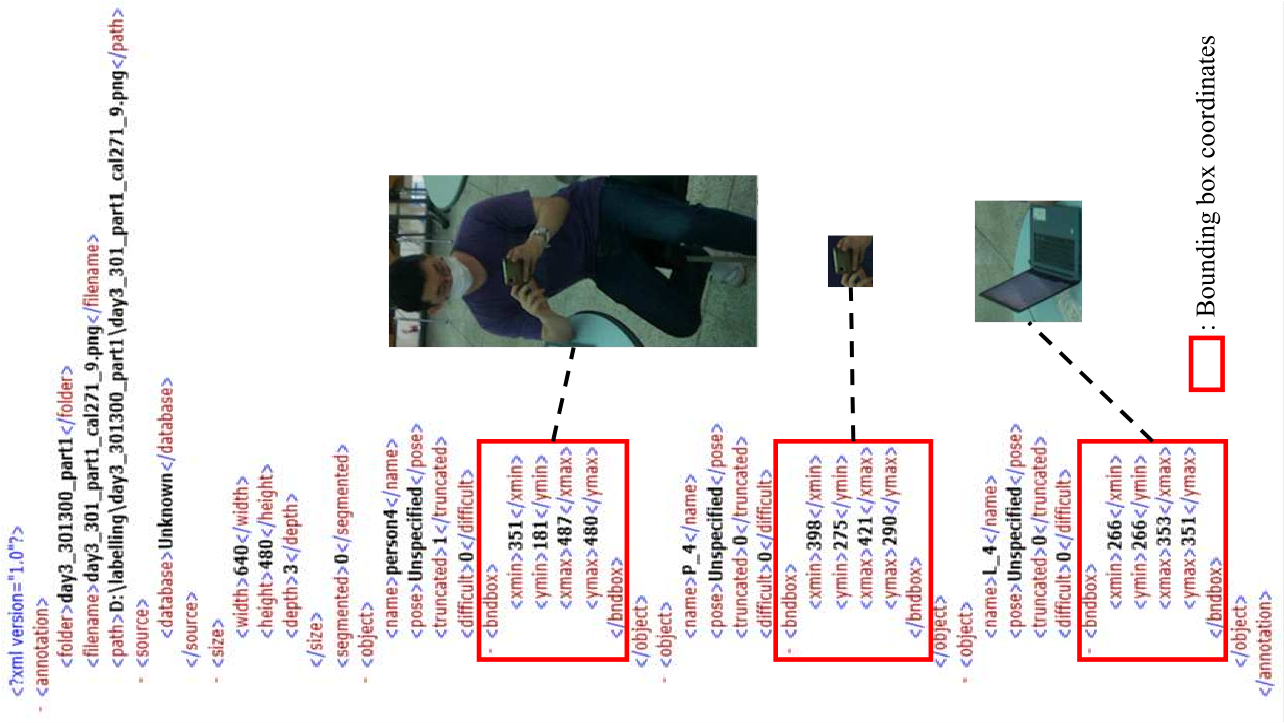}}
	\caption{Illustration of (a) each pair of RGB and depth images in VOMTC and (b) each data label.}
	\label{fig:eachVOMTCsample}
\end{figure}
\begin{figure*}[t]
\centering
{\includegraphics[width=2.0\columnwidth, height=7.0cm]{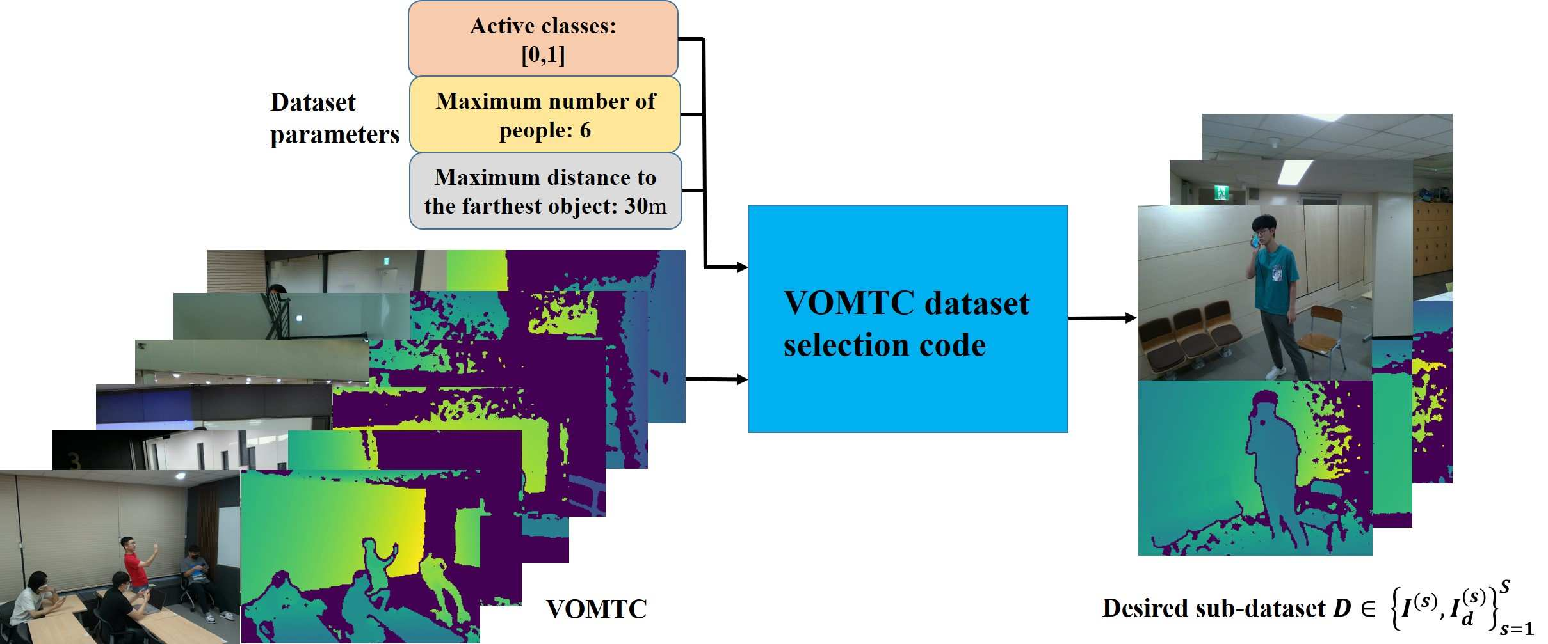}}
	\caption{Desired sub-dataset acquisition using the VOMTC dataset selection code. }
	\label{fig:datasetselectioncode}
 \vspace{-0.5cm}
\end{figure*}
We briefly summarize the notations used in this paper. We use uppercase boldface letters for matrices and lowercase boldface letters for vectors. 
Operations $(\cdot)^{T}$ and $(\cdot)^{H}$ denote transpose and conjugate transpose, respectively. Also, operators $\otimes$ and $\odot$ denote the Hadamard and Kronecker products, respectively. $\mathbb{C}$ and $\mathbb{R}$ denote the field of complex numbers and real numbers, respectively. $\lVert \cdot \rVert_{p}$ indicates the $p$-norm. $\mathbf{x}^{(i)}$ denotes the $i$-th column of matrix $\mathbf{X}$ and $x^{(i)}$ denotes the $i$-th entry of vector $ \mathbf{x}$. 
$\mathbf{C}_{true}(\mathcal{I}) = \left[ \mathbf{c}^{(1)}_{true}(\mathcal{I})\, \cdots\,\mathbf{c}^{(N)}_{true}(\mathcal{I}) \right]$, $\mathbf{C}_{ref}(\mathcal{I}) = \left[ \mathbf{c}^{(1)}_{ref}(\mathcal{I})\, \cdots\,\mathbf{c}^{(N)}_{ref}(\mathcal{I}) \right]$, and $\mathbf{C}(\mathcal{I}) = \left[ \mathbf{c}^{(1)}(\mathcal{I})\,\cdots \,\mathbf{c}^{(N)}(\mathcal{I}) \right]$ denote the matrices consisting of ground truth, fine-tuned, and baseline predicted class score vectors of objects in the given image $\mathcal{I} \in \mathbb{R}^{W\times H\times C}$, respectively.
Also, $\mathbf{c}_{true\,class}(\mathcal{I}) \in \mathbb{R}^{P}$, $\mathbf{c}_{ref\,class}(\mathcal{I}) \in \mathbb{R}^{P}$, and $\mathbf{c}_{class}(\mathcal{I}) \in \mathbb{R}^{P}$ denote the class score vectors representing ground truth, as well as the predicted probability values obtained from the fine-tuned model and the baseline model, respectively, for each object belonging to the given class.
Lastly, $\mathcal{I}$, $\mathcal{I}_{d}$, $\mathcal{I}_{c}$, and $\mathcal{I}_{c, d}$ denote whole RGB, whole depth, cropped RGB, and cropped depth images, respectively.

The rest of this paper is organized as follows. 
In Section II, we discuss the proposed VOMTC dataset in detail. 
In Section III, we provide an illustrative example demonstrating the training of a CV-aided object detector using VOMTC.
In Section IV, we introduce a few use cases of VOMTC.
In Section V, we present the simulation results to verify the performance gain of the proposed VBM trained by VOMTC and conclude the paper in Section VI.

\section{VOMTC dataset}\label{sec:3_intro}
In this section, we discuss the VOMTC dataset in detail.
We also compare VOMTC with conventional datasets.

\subsection{Characteristics and Elements of VOMTC dataset}\label{ssec:datasetdescription}
VOMTC is a large-scale vision dataset consisting of image samples captured from the RGB-depth camera\footnote{As for the RGB-depth camera, we use the Intel RealSense L515 RGB-depth camera.} attached to the BS.
The VOMTC dataset consists of 20,232 pairs of RGB and depth images collected from 2,083 distinct environments including convenience store, classroom, cafeteria, elevator, stairs, and subway station and at different times of the day (e.g., morning, afternoon, and night).
We manually annotate labels including the class (person, cell phone, or laptop) and the bounding box of objects appearing in each image of VOMTC.
We then divide VOMTC into three distinct sets: a training set, a validation set, and a test set, containing 17,299, 1,440, and 1,493 samples, respectively.

In VOMTC, we store the images and corresponding labels in two different folders named `image' and `label'.
In the `image' folder, there are three subfolders named `rgb', `depth', and `distance'.
The `rgb' subfolder contains RGB images, while the `depth' subfolder contains corresponding depth images (see Fig. 2(a)). 
In the `distance' subfolder, the distance to the point in each pixel obtained from the LiDAR sensor is included.
This information is available in JSON format [17]. 
On the other hand, the `label' folder provides labels for images of VOMTC in XML format [17].
Each label includes the width, height, and depth of the image, as well as objects labeled with one of the three classes and their bounding box coordinates.
Specifically, persons are labeled as `person', cell phones as `P', and laptops as `L'~(see Fig. 2(b)).

\begin{figure*}[t]
    \centering
    \includegraphics[width=2.0\columnwidth, height=8.0cm]{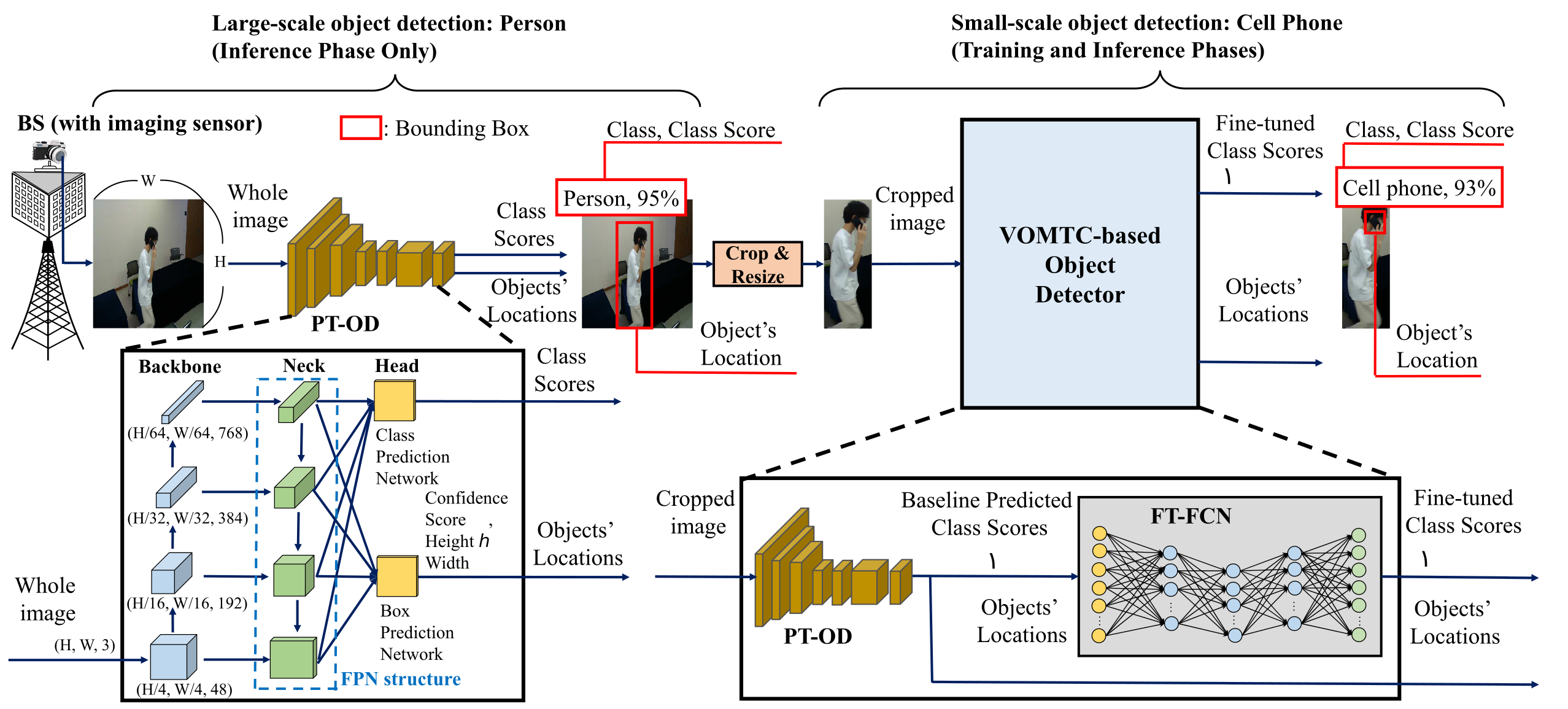}
	\caption{Overview and detailed network structure of the VOMTC-based object detector.}
  \vspace{-0.5cm}
\label{fig:proposedtechnique}
\end{figure*}
\subsection{Parameters of VOMTC dataset}\label{ssec:datasetparameters}

To help wireless researchers select the images and labels aligning with their specific needs, we design the dataset selection code. To execute the code, user needs to adjust the following key parameters:

\begin{itemize}
\item \textbf{Active classes} (defined by \texttt{activeClasses} in python code): In certain wireless tasks, user may only be interested in image samples containing a specific object class (e.g., laptops). The parameter \texttt{activeClasses} allows user to specify the desired classes appearing in each image. For example, by setting \texttt{activeClasses}$ =\left[ 1\right]$, one can collect depth-RGB image pairs consisting of cell phones.

\item \textbf{Maximum number of people} (defined by \texttt{maxnumPeople} in python code): 
This parameter allows user to choose the maximum number of people appearing in a VOMTC image sample.
For example, one can set \texttt{maxnumPeople} to $6$, to obtain depth-RGB image pairs containing $6$ or fewer people.

\item \textbf{Maximum distance to the farthest object} (defined by \texttt{maxDist} in python code): 
This parameter specifies the maximum distance between the object and the BS.
For example, one can set \texttt{maxDist} to $8$ to obtain vision samples where the distance to the object is less than $8$\,m.
\end{itemize}

By executing the code, user can obtain the desired sub-dataset $D \in \lbrace (\mathcal{I}^{(s)}, \mathcal{I}^{(s)}_{d})\rbrace^{S}_{s = 1}$ consisting of $S$ pairs of RGB and depth images (see Fig. 3).
For example, one can set \texttt{maxnumPeople} to $1$ or $5$ to collect images suited for the single-user MIMO (SU-MIMO) or multiple-user MIMO (MU-MIMO) systems, respectively.
If one wants the samples representing the far-field region exclusively, one can set \texttt{maxDist} to a large value (say, $30\,$m). 
Whereas, to choose the samples representing the near-field region where the array steering vector of the channel is affected by both the angle and the distance between the BS and the UE [18], one can set \texttt{maxDist} to a small value (say, $2\,$m).
\vspace{-0.2cm}

\begin{figure*}[t]
\centering
{\includegraphics[width=2.2\columnwidth, height=8.0cm]{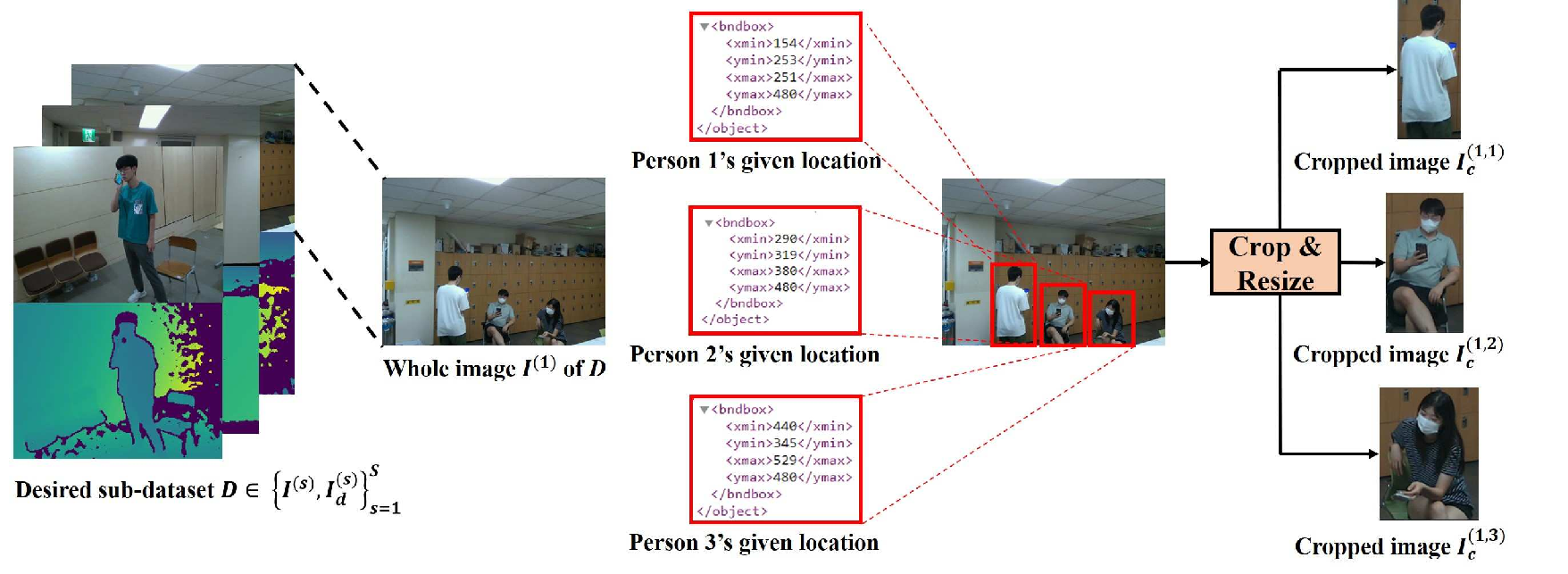}}
	\caption{Cropped images generation. }
	\label{fig:inputgeneration}
 \vspace{-0.5cm}
\end{figure*}
\subsection{Comparison with Other Datasets}\label{ssec:comparison}
In this subsection, we compare the proposed VOMTC and conventionally used datasets.
Here we briefly summarize the well-known image datasets:
\begin{itemize}

\item \emph{LSUN} [10] is a scene classification dataset featuring a large number of images for each scene category in its training set. 
In LSUN, there are 10 scene categories (e.g., dining room, bedroom, tower, etc.) and 20 object categories (e.g., chair, cow, table, etc.).

\item \emph{MS-COCO 2017} [19] is a large-scale dataset for the object detection, segmentation, and image captioning. It consists of 80 classes of objects and 164,000 images, with each image paired with 5 sentences.

\item \emph{ImageNet-1K} [20] is a widely used subset of ImageNet, a massive vision dataset consisting of more than 14 million images and spanning 1,000 object classes.

\item \emph{PASCAL VOC 2012} [21] is a popular image segmentation benchmark dataset containing 20 foreground classes and one background class. 
Since its original training set (consisting of 1,464 images) is relatively small, an augmented training dataset that consists of 10,582 images [47] is widely used [46].
 
\item \emph{CIFAR-10} [22] is another image dataset consisting of 60,000 32x32 color images of 10 classes (i.e., airplane, automobile, bird, cat, frog, horse, ship, and truck), with 6,000 images per class. 

\end{itemize}
Since the datasets we just mentioned do not contain wireless objects such as the cell phone or laptop, they might not be useful to train the CV-aided wireless tasks.
Recently, various datasets for the DL and CV-based wireless tasks have been proposed:
\begin{itemize}
 \item \emph{Vision-Wireless} (ViWi) [23] is a parametric and scalable dataset consisting of four sub-datasets for four outdoor MIMO-based scenarios with different camera distributions (`co-located' and `distributed') and views (`blocked' and `direct'). 
 \item \emph{ViWi Vision-Aided Millimeter-Wave Beam Tracking} (ViWi-BT) [24] is the second version of ViWi. ViWi-BT contains images captured by the co-located camera and mmWave MIMO beam codeword indices. 
\item \emph{Raymobtime} [25] is a synthetic dataset containing LIDAR, ray-tracing, GPS, matrix channel, and image data for mmWave MIMO vehicle-to-infrastructure systems.
It is suitable for the blockage prediction, power prediction, and angle estimation.
\item \emph{DeepMIMO} [26] is a synthetic dataset designed for mmWave/massive MIMO channels. Researchers can generate the dataset by selecting one of the given ray-tracing scenarios and then adjusting the parameters such as the number of active BSs, active UEs, number of BS antennas, antenna spacing, etc. 
\item \emph{ViWi blockage-prediction and object-detection datasets} [27] are datasets used for the blockage prediction and proactive beam/BS switch decision research. The former provides 54,000 multi-modal data samples in the form of 4-tuple: image, mmWave beam, link status, and position information and the latter is a small object detection dataset consisting of 600 images labeled with bounding boxes of cars, trucks, and buses.

\end{itemize}
Unlike the synthetic datasets that may result in a degraded performance in the test (i.e., real application) phase, VOMTC contains images acquired directly from the real world so that it contains real visual characteristics of wireless objects, including variations in specular, ambient, diffuse, and emissive light over time.
Furthermore, in contrast to the conventional wireless datasets containing many large-scale objects such as cars, trucks, and buses, VOMTC includes labels (i.e., bounding box, class, distance, angles) on the small-scale wireless objects (cell phones and laptops) so that it can be readily used for the millimeter and THz communication scenarios.
\vspace{-0.2cm}

\begin{figure*}[t]
\centering
{\includegraphics[width=2.0\columnwidth, height=6.5cm]{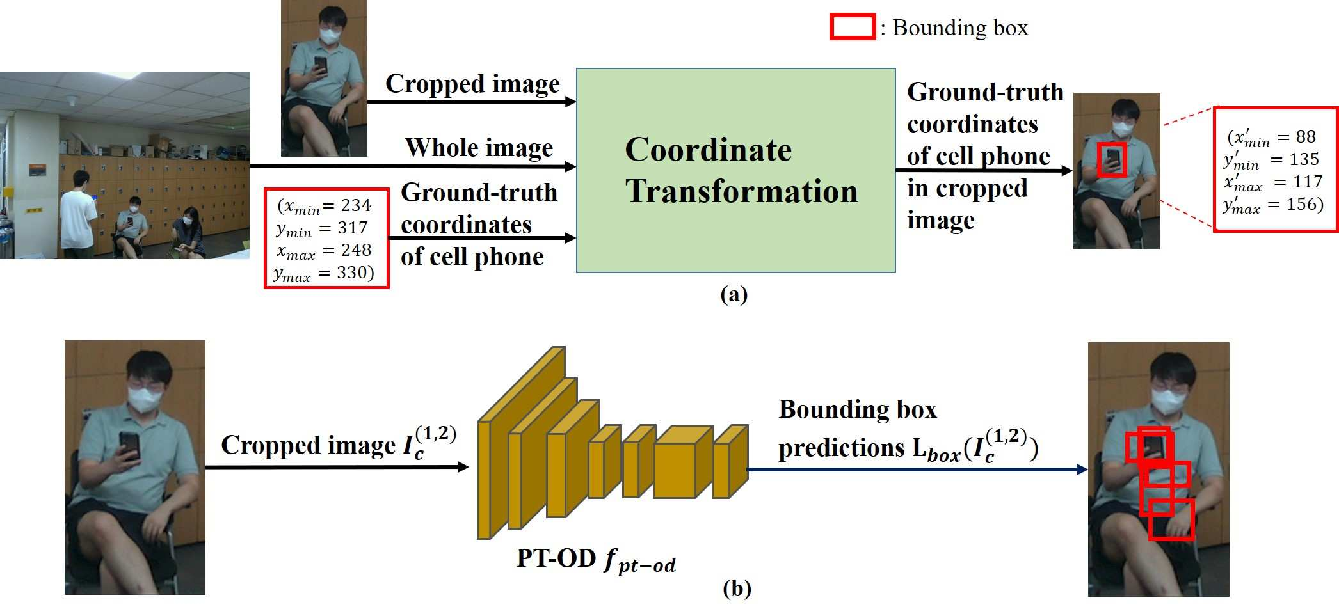}}
	\caption{Generation of (a) ground-truth bounding box coordinates of the cell phone in the cropped image and (b) bounding box predictions in the cropped image.}
  \vspace{-0.4cm}
	\label{fig:outputgeneration}
\end{figure*}
\section{Computer Vision-aided Beamforming using VOMTC}\label{sec:4}
In this section, we discuss two main steps (training and inference) of VOMTC-based VBM design.
\begin{itemize}
\item \emph{Training Phase of VBM}:
Training an object detector from the scratch requires many advanced graphic processing units (GPUs) and an enormous amount of training dataset.
To avoid such hassle, we perform the fine-tuning of the pre-trained object detector (PT-OD) using VOMTC.
In the fine-tuning phase, the VOMTC-based object detector, which consists of the PT-OD $f_{\text{pt-od}}$ and the fine-tuned fully connected network (FT-FCN) $f_{\text{ft-fcn}}$, learns to identify the cell phone in the cropped image $\mathcal{I}_{c}$ (i.e., the small-sized bounding box containing a person holding a cell phone). 
Specifically, we use $f_{\text{pt-od}}$ to generate class scores measuring the probability of each detected object belonging to a certain object class (e.g., cow, dog, and cat).
We then train $f_{\text{ft-fcn}}$ to generate the probability of each object belonging to the cell phone.
The overall process of the VOMTC-based object detector can be formulated as
\begin{align}
 \mathbf{L}_{box}(\mathcal{I}_{c}), \mathbf{C}(\mathcal{I}_{c}) = f_{\text{pt-od}}(\mathcal{I}_{c}) \label{eq:rgbprocessor}\\
\mathbf{C}_{ref}(\mathcal{I}_{c}) = f_{\text{ft-fcn}}(\mathbf{C}(\mathcal{I}_{c}); \Theta ) 
\label{eq:classscorerefiner}
\end{align}
where $\mathbf{L}_{box}(\mathcal{I}_{c})  \in \mathbb{R}^{Q \times 4}$ is the set of coordinates of four edges of $Q$ bounding boxes, 
$\Theta$ is the set of learnable network parameters of $f_{\text{ft-fcn}}$, and
$\mathbf{C}_{ref}(\mathcal{I}_{c}) =  \left[ \mathbf{c}_{ref\,{\text{phone}}}(\mathcal{I}_{c}) \,\mathbf{c}_{ref\,{\text{non-phone}}}(\mathcal{I}_{c}) \right] \in \mathbb{R}^{Q \times 2}$ and $\mathbf{C}(\mathcal{I}_{c}) = \left[\mathbf{c}_{ \text{phone}}(\mathcal{I}_{c}) \,\mathbf{c}_{ \text{person}}(\mathcal{I}_{c}) \right] \in \mathbb{R}^{Q \times 2}$ are fine-tuned and baseline predicted class scores of $Q$ objects in $\mathcal{I}_{c}$, respectively (see Fig. 4).
In the training phase, we fix the network parameters of $f_{\text{pt-od}}$ and focus on learning the optimal mapping function $f_{\text{ft-fcn}}^{*}$ minimizing the classification loss (the training process will be further described in Section III.B).

\item \emph{Inference Phase of VBM}: In the inference phase (i.e., application phase), VBM uses the VOMTC-trained object detector to perform the beamforming.
Specifically, we use the PT-OD $f_{\text{pt-od}}$ to identify a person holding a cell phone from the whole image $\mathcal{I}$\footnote{While cropped images are used in the fine-tuning phase, only whole images are available in the inference phase.
To obtain cropped images (i.e., inputs of the VOMTC-trained object detector), we adopt a two-stage object detection process.}. 
We next use the VOMTC-trained object detector to identify the cell phone from the cropped image $\mathcal{I}_{c}$ containing a person (see Fig. 4).
Using the acquired depth image and the identified 2D location of the cell phone in the RGB image, VBM derives the spherical coordinate representation of the cell phone's location $(r_w, \theta_w, \phi_w)$ (refer to Section IV.B) and then generates beams directing towards the acquired spherical coordinates (the beam generation process will be further described in Section III.C).
\end{itemize}
In the following subsections, we explain the wireless device identification, object detector training process, and location-aware beam transmission/reception processes in detail.

\subsection{Wireless Device Identification Via PT-OD and FT-FCN}

One essential component of the VOMTC-based object detector is the PT-OD $f_{\text{pt-od}}$, which identifies various objects in the cropped image $\mathcal{I}_{c}$ based on the class scores.
Basically, $f_{\text{pt-od}}$ consists of the following three main components (see Fig. 4):
\begin{itemize}
   \item  \textit{Backbone}: The backbone hierarchcially extracts the spatially-correlated features (e.g., color, shape, face) in various scales from the input RGB image. For example, global features (e.g., face and wall) are extracted at the top of the backbone and the local features (e.g., edge and curve) are extracted at the bottom.
    \item \textit{Neck}: Once the features are extracted at the backbone, the neck aggregates the extracted features in different scales. Note that the local feautres contain rich geometric information useful for the bounding box identification, while the global features contain rich semantic information. To achieve the robust performance in both localization and classification, the multi-scale features are aggregated using a feature pyramid network (FPN) structure, where the local features extracted at the top of backbone are delivered all the way to the bottom.
    \item \textit{Head}: Using the aggregated features as inputs, the head performs the bounding box detection and the class identification. To do so, the box prediction and the class prediction networks are needed in the head. First, the box prediction network computes the height and width of the object, along with the confidence score, which indicates the likelihood of the point being the center of the object. Second, the class prediction network computes the class score measuring the probability of each detected object belonging to a certain class, generating a class score matrix $\mathbf{C}(\mathcal{I}_{c}) \in \mathbb{R}^{Q \times J}$ where $Q$ and $J$ are the number of identified bounding boxes in $\mathcal{I}_{c}$ and the number of object classes in the PT-OD, respectively.
\end{itemize}

Using the selected information from $\mathbf{C}(\mathcal{I}_{c})$, the FT-FCN $f_{\text{ft-fcn}}$ learns to generate the probability of each object belonging to the cell phone.
Specifically, we first select the class score vectors belonging to the cell phone and person (i.e., $\mathbf{c}_{ \text{phone}}(\mathcal{I}_{c})$ and $\mathbf{c}_{ \text{person}}(\mathcal{I}_{c}))$ from $\mathbf{C}(\mathcal{I}_{c})$.
We then concatenate these vectors to generate the input vector $\mathbf{c}(\mathcal{I}_{c}) \in \mathbb{R}^{2Q}$ of $f_{\text{ft-fcn}}$.
After passing through multiple hidden layers of $f_{\text{ft-fcn}}$, $\mathbf{c}(\mathcal{I}_{c})$ is transformed into the output vector $\mathbf{c}_{ref}(\mathcal{I}_{c}) \in \mathbb{R}^{2Q}$, which is then reshaped into the matrix form $\mathbf{C}_{ref}(\mathcal{I}_{c}) \in \mathbb{R}^{Q \times 2}$.
\vspace{-0.3cm}

\begin{figure*}[t]
\centering
\vspace{-0.8cm}
\includegraphics[angle=270,origin=c, width=2.0\columnwidth, height= 10cm]{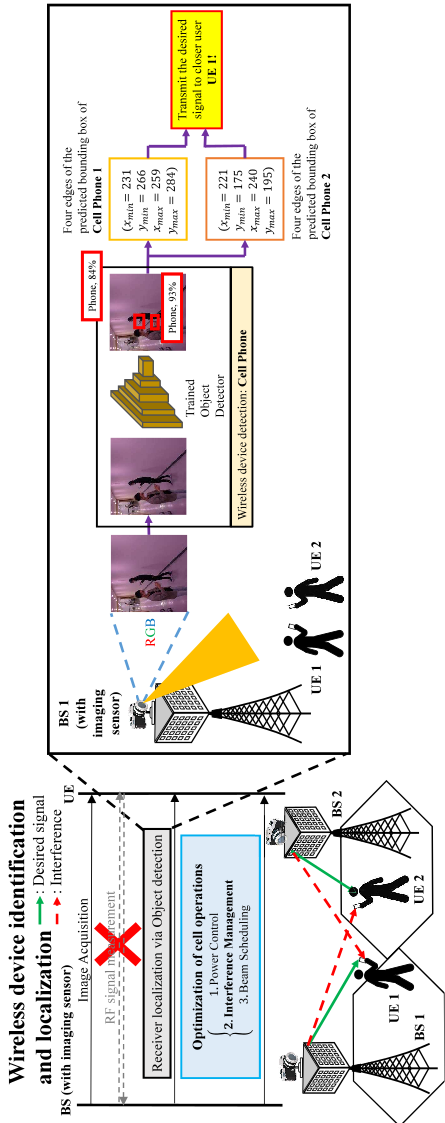}
\vspace{-4.4cm}
	\caption{Use Case 1: wireless device identification and localization.}
	\label{fig:usecase1}
 \vspace{-0.5cm}
\end{figure*}
\subsection{VOMTC-based Object Detector Training}

In this subsection, we briefly discuss the training label design for the VOMTC-based object detector.
Since the FT-FCN in the VOMTC-based object detector uses cropped images as training inputs, it requires new training labels (e.g., new bounding box in the cropped image). To generate these input-output pairs using VOMTC, we develop a training label design process, which can be performed by executing the training label design code\footnote{Check out https://github.com/islab-github/VOMTC.} consisting of the following steps:

\begin{enumerate}

\item For a given RGB image $\mathcal{I}^{(s)}$ in the desired sub-dataset $D$, we identify $Z$ bounding boxes of persons holding cell phones.
We then crop and resize these boxes to obtain $Z$ cropped RGB images $\lbrace \mathcal{I}^{(s, z)}_{c} \rbrace^{Z}_{z = 1}$, which are used as training inputs (see Fig. 5).

\item We obtain the four edges of the bounding box of the cell phone $(x^{'}_{\text{min}},y^{'}_{\text{min}}, x^{'}_{\text{max}}, y^{'}_{\text{max}})$ via simple coordinate transformation from $(x_{\text{min}},y_{\text{min}}, x_{\text{max}}, y_{\text{max}})$ in $\mathcal{I}^{(s)}$ to $(x^{'}_{\text{min}},y^{'}_{\text{min}}, x^{'}_{\text{max}}, y^{'}_{\text{max}})$ in $\mathcal{I}^{(s,z)}_{c}$ (see Fig. 6(a)). 

\item 
We feed each cropped image $\mathcal{I}^{(s,z)}_{c}$ into the PT-OD to obtain $Q$ predicted bounding boxes (see Fig. 6(b)).
Using these boxes and $(x^{'}_{\text{min}},y^{'}_{\text{min}}, x^{'}_{\text{max}}, y^{'}_{\text{max}})$, one can readily obtain the ground-truth class score matrix  $\mathbf{C}_{true}(\mathcal{I}_{c}) \in \mathbb{R}^{Q \times 2}$ needed for training $f_{\text{ft-fcn}}$.

\end{enumerate}

\subsection{Location-aware Beam Transmission and Reception}

Once the spherical coordinates of $K$ cell phones (i.e., small UEs) $[ (\theta^{(1)}_{w}, \phi^{(1)}_{w}) \, \cdots \, (\theta^{(K)}_{w}, \phi^{(K)}_{w})]$ are acquired, the BS can transmit sharp beams heading toward the extracted locations, as in (7) and (8).
Due to the low penetration power and severe path loss of the mmWave/THz signals, the scattering of signal is negligible so the LoS path becomes the dominant means of propagation. 
Thus the channel matrix $\mathbf{H}_k \in \mathbb{C}^{N \times M}$ from the BS to the $k$-th UE is expressed as
\begin{align}
    \mathbf{H}_k = \sqrt{\beta_k} \alpha_{k} \mathbf{a}_{M}(\theta^{(k)}_{\text{AoD}, w}, \phi^{(k)}_{\text{AoD}, w}) \mathbf{a}^H_{N}(\theta^{(k)}_{\text{AoA}, w}, \phi^{(k)}_{\text{AoA}, w}),
    \label{eq:downlinkchannel}
\end{align}
where $\beta_k \in \mathbb{R}_{+}$ is the large-scale fading coefficient, $\alpha_k \sim \mathcal{CN}(0,1)$ is the small-scale fading coefficient, and $\theta^{(k)}_{\text{AoD}, w} = \theta^{(k)}_{w}$ and $\phi^{(k)}_{\text{AoD}, w} = \phi^{(k)}_{w}$ are the AoD azimuth and elevation angles of the $k$-th UE, respectively.

We mention that VBM can find the AoD of the UE but the AoA at the UE is still unknown.
To find the azimuth and elevation of AoA ($\theta^{(k)}_{\text{AoA}, w}$ and $\phi^{(k)}_{\text{AoA}, w}$), we can employ the MUSIC algorithm [28].
By using a proper optimization technique, MUSIC easily determines $\theta^{(k)}_{\text{AoA}, w}$ and $\phi^{(k)}_{\text{AoA}, w}$ as components maximizing the MUSIC function $f_{\text{MUSIC}}(\theta^{(k)}_{\text{AoD}, w}, \phi^{(k)}_{\text{AoD}, w}, \theta^{(k)}_{\text{AoA}, w}, \phi^{(k)}_{\text{AoA}, w})$.
VBM then performs the beamforming using the array steering vector of the UE generated based on $\theta^{(k)}_{\text{AoA}, w}$ and $\phi^{(k)}_{\text{AoA}, w}$, which can be expressed as
\begin{align}
    \mathbf{a}_{N}(\theta^{(k)}_{\text{AoA}, w}, \phi^{(k)}_{\text{AoA}, w}) = \frac{1}{\sqrt{N}}[1\, \cdots e^{-j(N_{x}-1)\pi \theta^{(k)}_{\text{AoA}, w}}]^{T} \nonumber \\
    \otimes [1\, \cdots e^{-j(N_{y}-1)\pi \phi^{(k)}_{\text{AoA}, w}}]^{T}.
    \label{eq:VBMarraysteering2}
\end{align}
When the LoS path is blocked, we first quickly perform blockage detection using VBM and then use the conventional RF transmission of control/pilot signals to acquire the AoAs/AoDs associated with the paths between the BS and the UE.

In our work, we assume that the BS and the UE are equipped $M$ and $N$ analog phase shifters, respectively.
Due to the transmit power constraint of BS and UE, the analog precoder $\mathbf{f}_{k} = \mathbf{a}_{M}(\theta^{(k)}_{\text{AoD}, w}, \phi^{(k)}_{\text{AoD}, w}) \in \mathbb{C}^{M \times 1}$ and combiner $\mathbf{w}_{k} =  \mathbf{a}_{N}(\theta^{(k)}_{\text{AoA}, w}, \phi^{(k)}_{\text{AoA}, w}) \in \mathbb{C}^{N \times 1}$ are normalized as $\vert \vert  \mathbf{f} \vert \vert^2_2 = \vert \vert \mathbf{w} \vert \vert^2_2  =  1$.
In this setup, the received signal $y_k \in \mathbb{C}$ of the $k$-th UE is given by
\begin{align}
    y_k &= \sqrt{P_k} \mathbf{w}_k^H \mathbf{H}_k \mathbf{f}_k s_k + n_k, \ \ \ k = 1,\cdots, K, \label{eq:rxsignal}
\end{align}
where $P_k \in \mathbb{R}_{+}$ is the transmission power of the BS to the $k$-th UE, $s_k \in \mathbb{C}$ is the data symbol intended for the UE such that $\mathbb{E}[\vert s_k \vert^2] = 1$, and $n_k\sim \mathcal{CN}(0,\sigma_n^2)$ is the additive Gaussian noise.
Then the total downlink data rate $R_{\text{total}}$ between the BS and the $K$ UEs can be expressed as
\begin{align}
    R_{\text{total}} &= \sum^{K}_{k = 1} \log_2 \left( 1 + \frac{P_k \vert \mathbf{w}_k^H \mathbf{H}_k \mathbf{f}_k\vert^2 }{\sigma_n^2} \right), \ \text{for} \ k=1,\cdots, K. \label{eq:rate}
\end{align}
By using the VOMTC-trained object detector to identify UEs in an image, the BS can readily obtain their exact locations, generating the beamforming vector aligned with the channel $\mathbf{H}_{k}$.

It is also worth mentioning that the channel vectors of UEs are asymptotically orthogonal to each other when the number of antennas is sufficiently large [29], which is true in the upcoming 6G systems where 
hundreds or thousands of antennas are employed (i.e., IRS, holographic MIMO, and extreme MIMO (xMIMO)).
This means that the multi-user interference, caused by the beams heading toward the positions of other UEs, might be negligible.


\section{Use Cases of VOMTC}
In this section, we discuss four use cases of VOMTC: wireless device identification/localization, mmWave/THz beamforming, wireless device blockage detection/mobility management, and intelligent reflecting surface (IRS) phase shift control.

\subsection{Wireless Device Identification and Localization}\label{sec:2_1}

\begin{figure*}[t]
\centering
\vspace{-0.8cm}
\includegraphics[angle=270,origin=c, width=2.0\columnwidth, height= 10cm]{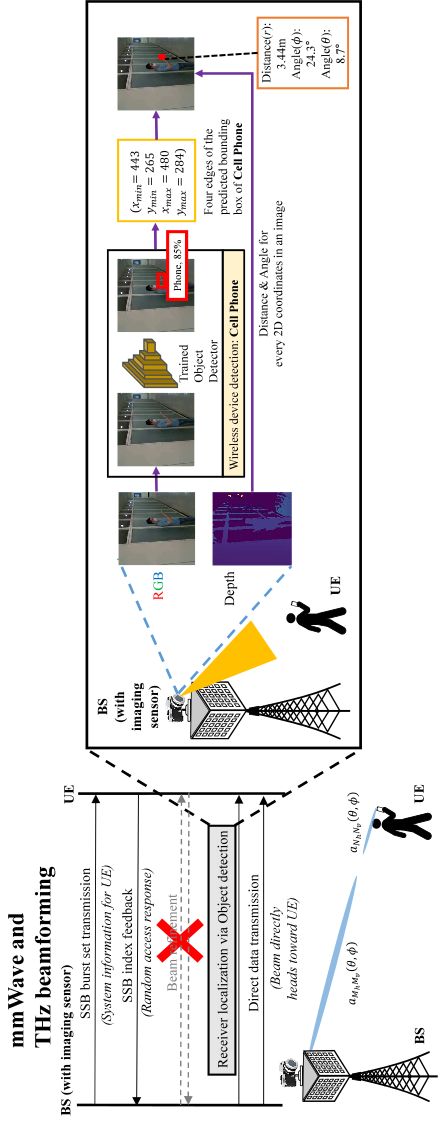}
	\vspace{-4.4cm}
	\caption{Use Case 2: mmWave and THz beamforming.}
	\label{fig:usecase2}
  \vspace{-0.5cm}
\end{figure*}


Recently, location awareness has received the special attention due to the wide range of add-on applications such as location-based advertisement and emergency services [30]-[32].
Also, accurate localization of the wireless device is crucial for the optimization of physical layer applications such as the beamforming and precoding,  power control, and interference management [33].

Over the years, triangulation-based positioning techniques leveraging RF signal information such as time of arrival (ToA), received signal strength indication (RSSI), and angle of arrival (AoA) have been widely used [34], [35].
In these schemes, 3D position $(x, y, z)$ of the target wireless device is extracted by identifying the intersection of the spheres centered at three or more anchor nodes with the radius of the measured distances.
Major weakness of conventional positioning techniques is that when the signal propagates through non-line-of-sight (NLoS) paths, the measured distance will be much longer than the actual distance, resulting in a severe localization error.
Also, since these techniques require multiple signal measurements from the distinct anchor nodes, they might not perform well if any of these measurements are missing or target/source is moving.

To estimate the precise location without relying on RF signals, one can use the visual sensor along with the CV technique.
Essence of this approach is to use the object detector to identify the class of the target object and its 2D location $(x, y)$ in an image [36].
In designing an object detector specialized for the accurate identification and localization of wireless devices, the VOMTC training set will be of great help.
Unlike conventional localization techniques requiring multiple signal measurements, the VOMTC-based object detector can identify the location of a target wireless device using an image captured by a camera attached to the BS (see Fig.~\ref{fig:usecase1}).
Noting that the sensing information (i.e., images) are obtained through visible light (400$\sim$790 THz), NLoS signal components can be naturally suppressed.

\subsection{mmWave and Terahertz beamforming}\label{sec:2_2}

One major bottleneck of the mmWave/THz communications is the
severe attenuation of signal power caused by high diffraction, penetration losses, and atmospheric absorption. 
To compensate for the severe path loss, a beamforming technique using the massive MIMO systems has been popularly used [37]-[40].
To maximize the beamforming gain, the beamforming vector needs to be aligned with the array steering vectors of signal propagation paths.
To do so, the BS should have the precise downlink channel information.
In 5G NR, a set of analog beam codeword called a beam codebook is used for the channel acquisition. 
A well-known drawback of the 5G NR beamforming is the mismatch between the pre-defined discrete beam direction and the real direction.
For example, in the 5G NR system where $1024$  beams are used, each beam covers the circle sector of $8$ square degrees on average so that the beam direction error in the worst case will be $4$ degrees in azimuth and elevation angles, ending up with more than 20\% reduction in beamforming gain [41].
Another drawback of 5G NR beamforming is the considerable latency caused by the beam sweeping and beam refinement operations.
In the beam refinement process, it takes $30\,$ms to transmit $4$ channel state information-reference signal (CSI-RS) beams.
Due to the short coherence time of mmWave/THz channel (e.g., $9\,$ms when the mobile is moving at $30\,$km/h speed), the beam direction
will be completely misaligned even with a small movement of the mobile, resulting in a degradation of beamforming gain [3].

CV-aided beamforming can be an appealing solution to the problem at hand.
The key characteristic of CV-aided beamforming is that it extracts the location of a wireless device from the information obtained by imaging sensors (e.g., RGB camera, LiDAR, and radar) and then transmits the non-discretized (continuous) beam heading toward the identified location. 
Specifically, CV-aided beamforming uses an object detector to extract the device's 2D coordinates $(x,y)$ from the captured image (see Section IV for details).
Then, using the depth image obtained from the LiDAR sensor, the distance to each point in every pixel is obtained.
Using the acquired depth information $r$ and the 2D location $(x, y)$ of the target device, one can obtain the 3D cartesian coordinates $(x,y,z)$ ($z = \sqrt{r^2 - (x^2 + y^2)}$), which can be converted in the form of 3D spherical coordinates $(r_{w}, \theta_{w}, \phi_{w})$ using $\theta_{w} =\arctan\frac{\sqrt{x^2 + y^2}}{z}$ and $\phi_{w} =\arctan{\frac{y}{x}}$.

Once $(r_{w}, \theta_{w}, \phi_{w})$ is obtained, the BS generates the beamforming vector heading toward the target wireless device (see Fig.~\ref{fig:usecase2}).
For instance, in the downlink THz MIMO system where the BS equipped with $M=M_x \times M_y$ uniform planar array (UPA) serves the target device with $N=N_x \times N_y$ antenna elements, the analog beams can be expressed as the Kronecker products of azimuth and elevation array steering vectors:
\begin{align}
    \mathbf{a}_{M}(\theta_w, \phi_w) = \mathbf{u}_{M_x}(\theta_w) \otimes \mathbf{u}_{M_y}(\phi_w) \label{eq:downlinkarraysteering1}\\
    \mathbf{a}_{N}(\theta_w, \phi_w) = \mathbf{u}_{N_x}(\theta_w) \otimes \mathbf{u}_{N_y}(\phi_w)
    \label{eq:downlinkarraysteering2}
\end{align}
where $\mathbf{u}_{L}(\theta) = \frac{1}{\sqrt{L}} [1\,e^{j\pi \theta} \cdots e^{j(L-1)\pi \theta}]^{T}$ is the array steering vector for the uniform linear array (ULA) with $L$ antennas.
Since BS can directly identify the beam direction without the codebook quantization, CV-aided beamforming can achieve a substantial improvement in beamforming gain.

\subsection{Wireless Device Blockage Detection and Mobility Management}\label{sec:2_3}
\begin{figure*}[t]
\centering
\vspace{-0.8cm}
\includegraphics[angle=270,origin=c, width=2.0\columnwidth, height= 10cm]{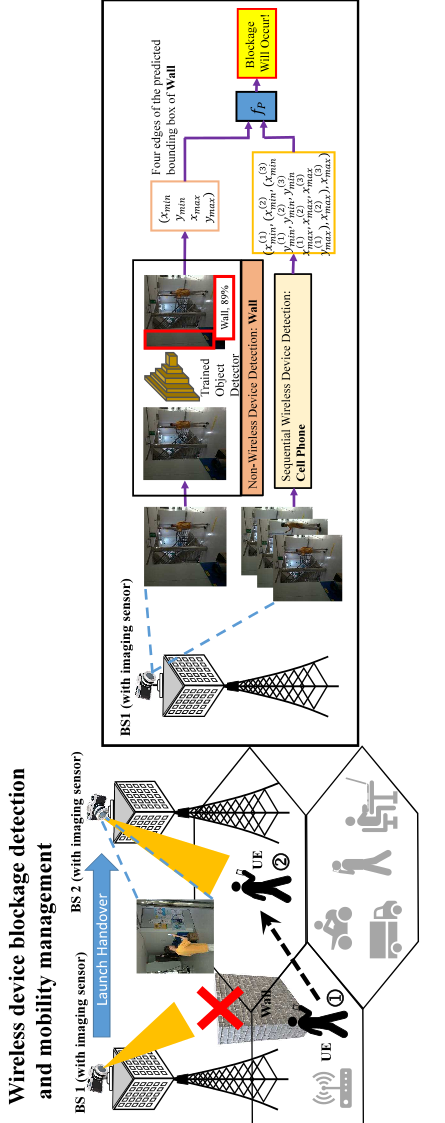}
	\vspace{-4.4cm}
	\caption{Use Case 3: wireless device blockage detection and mobility management.}
	\label{fig:usecase3}
  \vspace{-0.5cm}
\end{figure*}
Due to the high directivity and severe path loss, reflection and scattering of the signal will be weak in the THz band, leaving the line-of-sight (LoS) path as the dominant path of signal propagation. 
Indeed, the Rician K-factor, a ratio of the power of the strongest path over the sum of powers of other paths, is around 20 dB in 0.4 THz band [42], meaning that the received signal power of LoS path is almost 100 times higher than that of non-line-of-sight (NLoS) path.
For this reason, it is of importance to detect sudden blockages in the LoS path to ensure seamless connectivity [4]. 
In 4G LTE and 5G NR, the signal blockage is detected when the received power of the reference signals (RSs) falls below a pre-defined threshold [43].
Since this reactive power measurement process requires a considerable transmission delay (e.g., $30\,$ms to transmit $4$ CSI-RS beams [3]), latency in the blockage detection cannot be avoidable, causing a frequent radio link failure (RLF).

DL-based CV can readily alleviate this problem by quickly checking whether the target wireless device exists (no blockage) or not (blockage occurs) in the captured image.
The VOMTC dataset is a good fit for this task.
For instance, one can find out the positions of the $K$ non-wireless objects (e.g., wall, door, and table) causing LoS blockages $\bm{\delta} = [(x^{(1)}, y^{(1)}) \cdots (x^{(K)}, y^{(K)})]^T$ and the $S$ sequential positions of the cell phone $\mathbf{p} = [(x^{(1)}, y^{(1)}) \cdots (x^{(S)}, y^{(S)})]^T$ via VOMTC-trained object detectors.
Using $\mathbf{p}$ and $\bm{\delta}$ as inputs to the DNN model predicting the occurrence of the cell phone blockage, one can quickly and accurately initiate the handover process~(see Fig.~\ref{fig:usecase3}). 
The mapping function $f_P$ that predicts the LoS blockage from $\mathbf{p}$ and $\bm{\delta}$ can be expressed as
\begin{align}
    b = f_P (\mathbf{p}, \bm{\delta} ; \boldsymbol{\omega}),
\end{align}
where $b$ is the binary classifier indicating whether the blockage will occur ($b = 1$) or not ($b = 0$) and $\boldsymbol{\omega}$ is the set of the DNN parameters.

\subsection{Intelligent Reflecting Surface (IRS) Phase Shift Control}\label{sec:2_4}
\begin{figure*}[t]
\centering
\vspace{-0.8cm}
\includegraphics[angle=270,origin=c, width=2.0\columnwidth, height= 10cm]{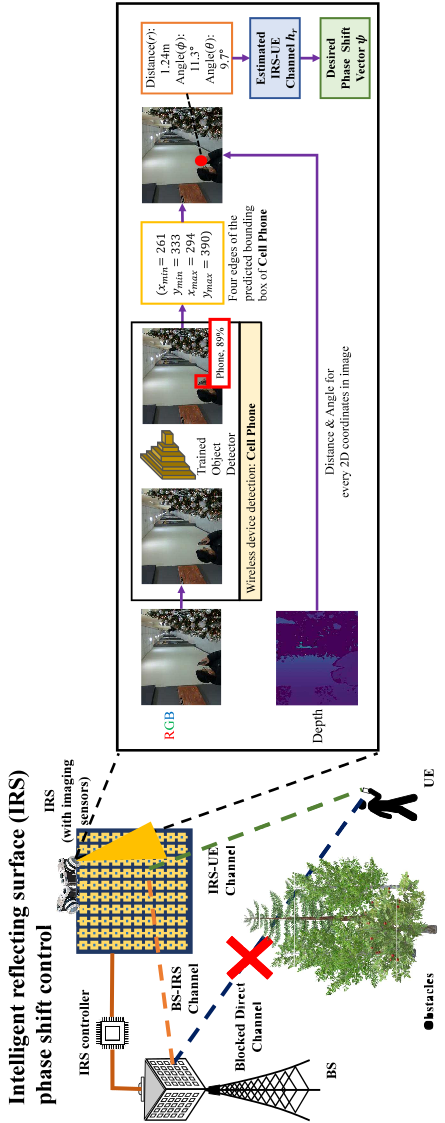}
	\vspace{-4.4cm}
	\caption{Use Case 4: intelligent reflecting surface (IRS) phase shift control.}
	\label{fig:usecase4}
  \vspace{-0.5cm}
\end{figure*}
Intelligent reflecting surface (IRS) has emerged as a promising solution to establish a virtual LoS link between a wireless device and the BS when the direct path is blocke [44].
In order to maximize the throughput gain of the IRS-aided systems, the phase of IRS needs to be adjusted such that the incident signals are reflected to the target destination accurately.
To do so, the BS needs to acquire not only the direct channel between the BS and UE but also the IRS-reflected channels. 
However, this task is challenging due to the huge pilot overhead and channel estimation error induced by a large number of reflecting elements $N$ (e.g., $N = 64 \sim 1024$) [45].
Since IRS has no dedicated RF chains to transmit or receive the pilot signals, the BS needs to estimate the full-dimensional IRS reflected channels with limited pilot signals, which introduces considerable channel estimation errors [45].

To determine the appropriate IRS phase shifts without relying on the channel estimation at the BS side, one can use the VOMTC-trained object detector.
This object detector extracts the location of a mobile from the sensing image, using which the geometric channel parameters of the IRS-UE channel can be obtained.
Once the channel is reconstructed, the phase shift vector that maximizes the downlink data rate can be generated via a proper optimization technique~(see Fig.~\ref{fig:usecase4}). 
Specifically, the VOMTC-trained object detector can identify the visible UE and its 2D location $(x_{\text{UE}},y_{\text{UE}})$ in an image captured by the imaging sensors installed on the IRS.
The spherical coordinates of the UE $(\theta_{\text{UE}}, \phi_{\text{UE}})$ can then be obtained using the same procedure introduced in Section IV.B. 
From the acquired coordinates of the UE, the IRS-UE channel vector $\mathbf{h}_{r} \in \mathbb{C}^{N}$ can be obtained as
\begin{align}
\mathbf{h}_{r} &= \sqrt{\beta_{r}} \mathbf{a}_{\text{IRS}}(\theta_{\text{UE}}, \phi_{\text{UE}}),
\label{eq:downlinkchannel} 
\end{align}
where $\beta_{r}$ and $\mathbf{a}_{\text{IRS}}(\theta_{\text{UE}}, \phi_{\text{UE}}) = \mathbf{a}_{\text{IRS}, x}(\theta_{\text{UE}}, \phi_{\text{UE}}) \otimes 
 \mathbf{a}_{\text{IRS}, y}(\theta_{\text{UE}}, \phi_{\text{UE}}) \in \mathbb{C}^{N}$ are the path loss and the IRS array response vector, respectively.
Note that $\mathbf{a}_{\text{IRS}, x}(\theta_{\text{UE}}, \phi_{\text{UE}})$ and  $ \mathbf{a}_{\text{IRS}, y}(\theta_{\text{UE}}, \phi_{\text{UE}})$ are expressed as $[1 \cdots e^{-j \pi (N_{x}-1) \text{cos}\theta_{\text{UE}} \text{sin}\phi_{\text{UE}} }]^{T}$ and
$[1 \cdots e^{-j \pi (N_{y} - 1) \text{sin}\theta_{\text{UE}} \text{sin}\phi_{\text{UE}}}]^{T}$, respectively.
Once $\mathbf{h}_{r}$ is obtained, the desired IRS phase shift vector $\bm{\psi} = [e^{jw_{1}} \cdots e^{jw_{N}}]^{T}$ can be generated by solving the problem: 
\begin{align}
 \label{phaseshiftoptimization}
    \max_{\bm{\psi}}\, & \log_2 \left( 1+ \frac{ P || (\mathbf{h}_{r}\text{diag}(\bm{\psi}))\mathbf{x}||^2}{\sigma^2_0} \right) \\
     \text{s.t.}\, &|\psi_{n}| = 1, \quad \forall n = 1, 2, \cdots N, \label{constraint}
\end{align}
where $N$ is the number of reflecting elements of the IRS, $P$ is the transmit power of BS, $\mathbf{x}$ is the transmitted signal, and $\sigma^2_0$ is the noise variance. 


\section{Experiments and Discussions}
\vspace{0.2cm}

\subsection{Experiment Setup}\label{ssec:dataset}

As for the PT-OD $f_{\text{pt-od}}$ used in our two-stage approach, we employ EfficientDet-D8 [13], a state-of-the-art object detector outperforming the conventional detectors such as Faster R-CNN [11] and YOLO [12]. 
Specifically, we use the model pre-trained on the MS-COCO 2017 dataset, which consists of 80 classes of objects and 118,000 training images.
As for the FT-FCN $f_{\text{ft-fcn}}$ used in our two-stage approach, we use the DNN consisting of input, output, and 3 hidden layers.
In our experiments, we set the number of identified bounding boxes $Q$ to $200$ to consider only the bounding boxes having the top $200$ class scores in the captured image.
Since we use the class score vector $\mathbf{c}(\mathcal{I}_{c}) \in \mathbb{R}^{2Q}$ as the input of  $f_{\text{ft-fcn}}$ to generate the same-sized class score vector $\mathbf{c}_{ref}(\mathcal{I}_{c})$, we set both the number of input and output layers to $400$. 
We also set the number of three hidden layers to $300$, $200$, and $300$, respectively.
\begin{figure}[t]
	\centering
 \includegraphics[width=1.0\columnwidth, height = 3.5cm]{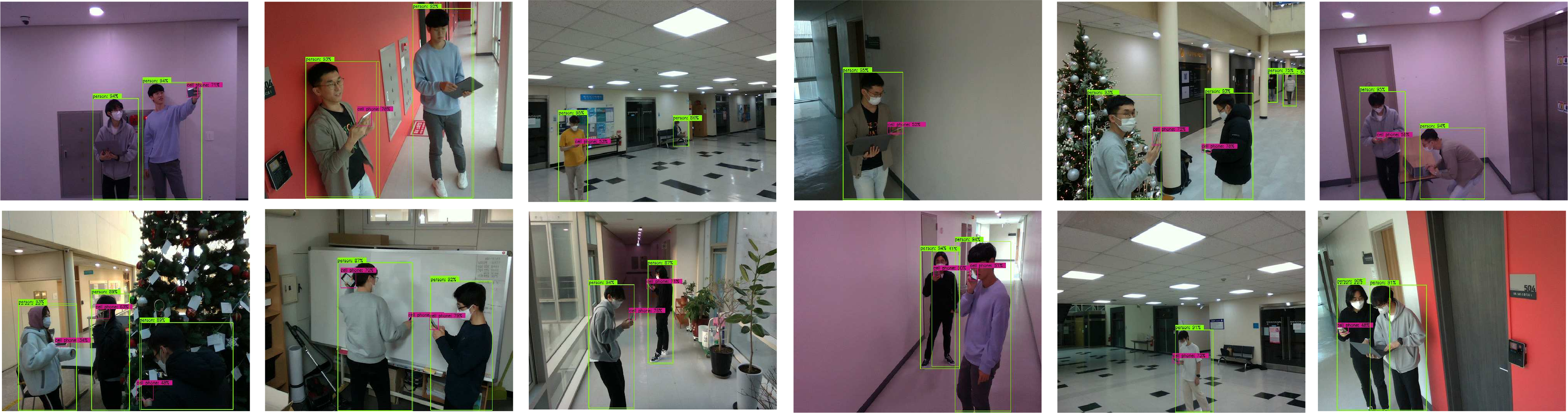}
	\vspace{-0.2cm}
	\caption{Object detection results of the VOMTC-trained object detector on the VOMTC test set. Green and red bounding boxes indicate persons and cell phones, respectively.}
	\label{fig:testresult}
 \vspace{-0.3cm}
\end{figure}
\begin{figure}[t]
	\centering
	\includegraphics[width=1.0\columnwidth, height = 3.5cm]{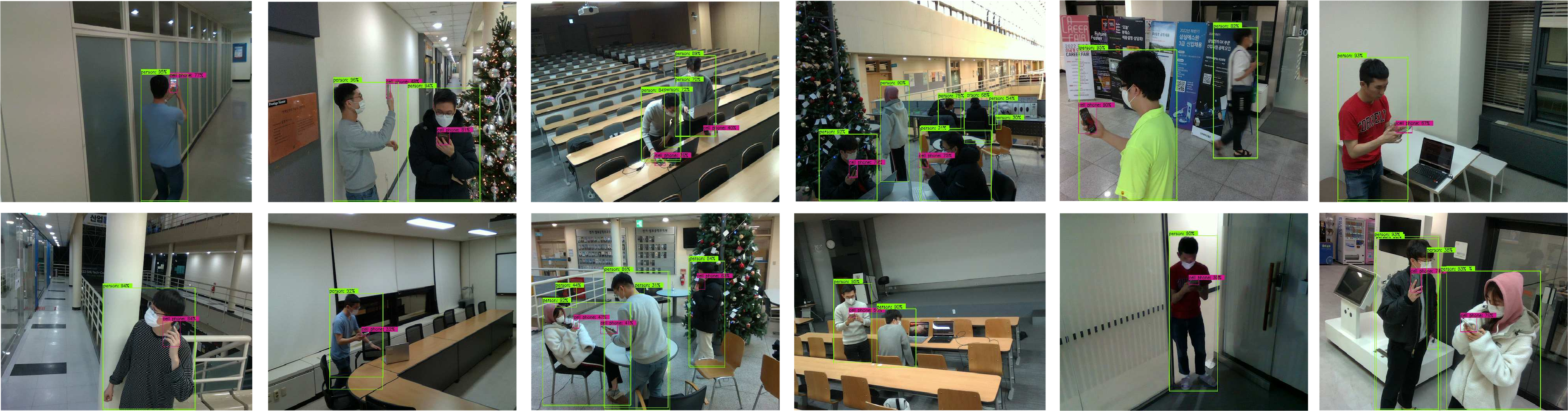}
	\vspace{-0.2cm}
	\caption{Object detection results of the VOMTC-trained object detector on the VOMTC validation set. Green and red bounding boxes indicate persons and cell phones, respectively.}
	\label{fig:valresult}
 \vspace{-0.3cm}
\end{figure}
\begin{table*}[t]
    \centering
    \caption{Object localization performance evaluated on the VOMTC Test Set.}
   
    \resizebox{2.0\columnwidth}{!}{
    \begin{tabular}{c|ccc|cc}
        \hline
        &
        \multicolumn{3}{c}{Cell phone} & \multicolumn{2}{|c}{Localization error}  \\
           & Precision (\%) & Recall (\%) & F1 score (\%) & Azimuth Angle (rad) & Elevation Angle (rad) \\
         \hline
        VOMTC-trained Object Detector & \textbf{94.46}    & \textbf{81.04}  & \textbf{87.24}   & \textbf{0.0804}  & \textbf{0.0804}        \\
        EfficientDet-D8  & 94.17   & 76.78  & 84.59 & 0.0972  & 0.0973     \\
         \hline
        \end{tabular}}
    \label{tab:vision_result}
\end{table*}
\vspace{0.4cm}
\begin{table*}[t]
    \centering
    \caption{Object localization performance  evaluated on the VOMTC Validation Set.}
    
    \resizebox{2.0\columnwidth}{!}{
    \begin{tabular}{c|ccc|cc}
        \hline
        &
        \multicolumn{3}{c}{Cell phone} & \multicolumn{2}{|c}{Localization error}  \\
           & Precision (\%) & Recall (\%) & F1 score (\%) & Azimuth Angle (rad) & Elevation Angle (rad) \\
         \hline
        VOMTC-trained Object Detector & \textbf{90.74}    & \textbf{70.70}  & \textbf{79.48}   & \textbf{0.1212}  & \textbf{0.1208}        \\
        EfficientDet-D8  & 90.10   & 66.70  & 76.66 & 0.1371 & 0.1365   \\
         \hline
        \end{tabular}}
    \label{tab:vision_val_result}
\end{table*}

In order to train $f_{\text{ft-fcn}}$, we first select the desired samples from VOMTC by specifying the three adjustable parameters as \texttt{activeClasses}$ = \left[ 0,1\right]$, \texttt{maxnumPeople}$ = 6$, and \texttt{maxDist}$ = 30$, and then running the dataset selection code.
Next, we run the training label design code, obtaining $9,302$ cropped RGB images (i.e., $\lbrace \mathcal{I}^{(s, z)}_{c} \rbrace^{9,302}_{z = 1}$) and their labels from the previously selected $7,654$ whole RGB images (i.e., $\lbrace \mathcal{I}^{(s)}\rbrace^{7,654}_{s = 1}$).
In the training phase, we use the AdamW optimizer, a well-known optimization tool to maintain the robustness of the learning process [48], with a learning rate $\eta$ of $10^{-4}$ and batch size $B$ of $32$.
Once $f_{\text{ft-fcn}}$ is trained, in the inference phase, we first use EfficientDet-D8 to detect a person and then use $f_{\text{ft-fcn}}$ to identify a cell phone.

In our experiments, we extensively evaluate the cell phone detection performance of the VOMTC-trained object detector on both VOMTC test and validation sets.
In Fig.~\ref{fig:testresult} and Fig.~\ref{fig:valresult}, we illustrate the detection results of the VOMTC-trained object detector on the VOMTC test and validation sets, respectively.
As performance metrics, we use precision, recall, and the F1 detection score [49], which are defined as $\frac{\text{correctly detected objects}}{\text{all target objects}}$, $
\frac{\text{correctly detected objects}}{\text{total detected objects}}$, and harmonic mean of precision and recall, respectively.
For comparison, we choose an object detection approach that uses EfficientDet-D8 in both large and small-scale object detection stages [3]\footnote{In the large and small-scale object detection stages, the precision/recall/F1 score for person and cell phone are obtained, respectively.}.

In the beam management simulations, we consider the THz MIMO system where the BS equipped with $M = 64$ transmit antennas serves the UE equipped with $N = 4$ receive antennas. 
The UE is randomly distributed in a square area of 20$\times$20$\,$m$^2$.
We set the carrier frequency $f_{c}$, transmission power $P$, noise variance $\sigma^2_{n}$, and bandwidth $W$ to 0.1$\,$THz, 2\,W, and 0.1, and 100$\,$MHz, respectively.
We also set the UE height to the average adult height (165cm) [50].
As the path loss, we use the indoor path loss model specified in 3GPP TR 38.901 Rel. 17\footnote{$\beta_k(r) = - \left( 31.84 + 21.5 \,\text{log}_{10} (r) + 19 \, \text{log}_{10} (f_c) \right)$, where $r$ is the distance between the BS and the UE.} [51].
For performance comparison, we use four baselines: 
\begin{enumerate} 
\item  CVBM [3]: In this scheme, the BS exploits EfficientDet-D8 to derive the 2D location of the UE in the RGB image.
The BS then generates a 3D beam directed towards the UE using the derived 2D location of the UE and the captured depth image.
\item  Codebook-based beamforming using object detection (OD) [52]: In this scheme, the BS picks the best beam codeword by identifying the codeword aligned to the UE's 2D location. Note that the UE's 2D location is derived using YOLOv3 and GPS.
\item Codebook-based beamforming using image classification (IC) [53]: In this scheme, the BS captures an image for each area covered by the given beam index and then selects the best beam codeword by identifying the image containing the UE.
\item  5G NR beam management (5G-BM) using $8$-bit DFT-based beam codebook with oversampling ratio of $4$ [54]: In this scheme, the BS selects the beam codeword that maximizes the reference signal received
power (RSRP) from the predefined codebook.

\end{enumerate}

\begin{figure*}[t]
\centering
\subfloat[]{\includegraphics[angle=0,origin=c,width=1.0\columnwidth, height = 5.0cm]{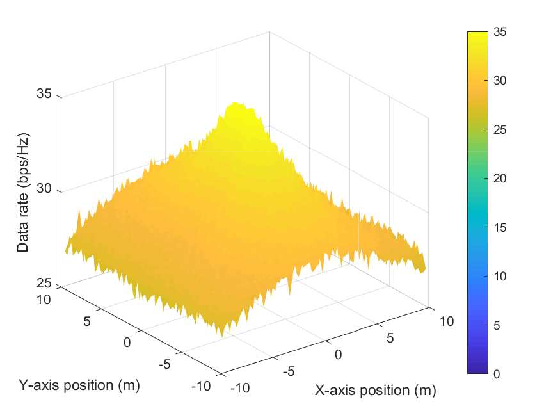}}\\
\subfloat[]{\includegraphics[angle=0,origin=c,width=1.0\columnwidth, height = 5.0cm]{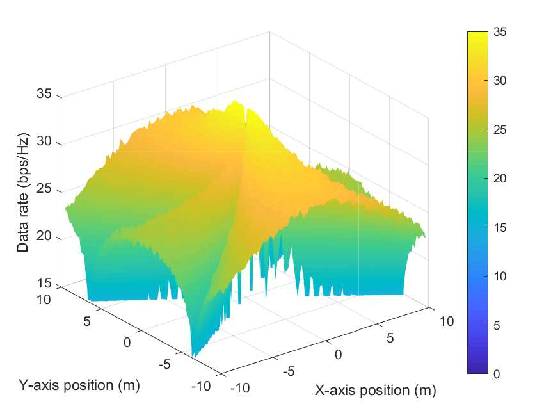}}
\subfloat[]{\includegraphics[angle=0, origin=c, width=1.0\columnwidth, height = 5.0cm]{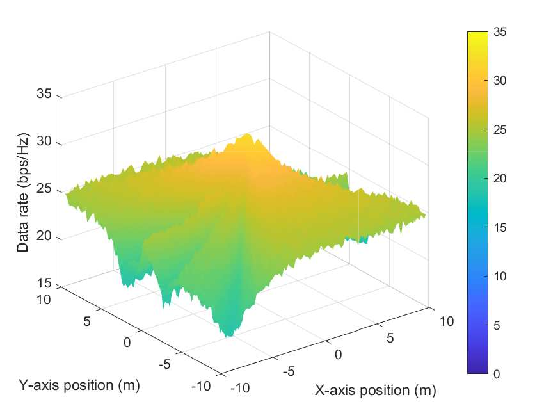}}\\
\subfloat[]{\includegraphics[angle=0,origin=c,width=1.0\columnwidth, height = 5.0cm]{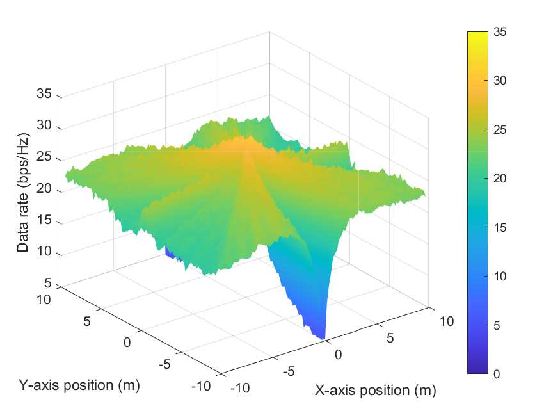}}
\subfloat[]{\includegraphics[angle=0, origin=c, width=1.0\columnwidth, height = 5.0cm]{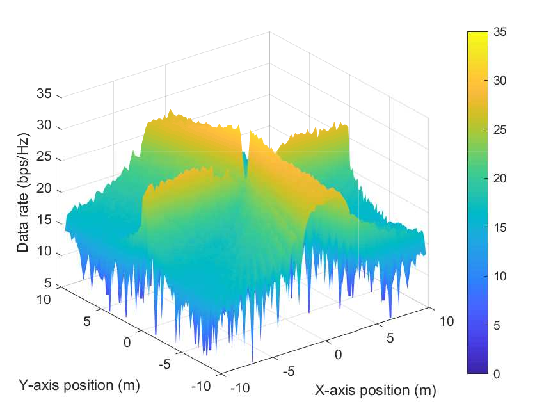}}
	\caption{Illustration of data rate at each user position (x,y) when the BS with the height of 3$\,$m located at (0,0) uses (a) VBM, (b) CVBM, and (c) Codebook-based beamforming using OD, (d) Codebook-based beamforming  using IC, and (e) 5G-BM.}
	\label{fig:throughputresult}
 \vspace{-0.4cm}
\end{figure*}

\subsection{Experiment Results}\label{ssec:vision_result}

Table~\ref{tab:vision_result} and Table~\ref{tab:vision_val_result} present the cell phone detection performance of the VOMTC-trained object detector and popularly used EfficientDet-D8.
We observe that the VOMTC-trained object detector outperforms conventional one in terms of precision, recall, and the F1 detection score. For instance, the VOMTC-trained object detector shows an average of 6\% improvement in the detection recall for the validation and test sets.

Based on the localization results, we evaluate the data rate of five beam management techniques as a function of user location.
In Fig. 13, we illustrate the achievable data rate of a cell phone using a 3D color map. 
We observe that VBM provides high throughput for most of the service area, achieving about 36\%, 29\%, 21\%, and 13\% improvements in the average data rate over 5G-BM, codebook-based beamforming using IC, codebook-based beamforming using OD, and CVBM, respectively.
We note that CVBM does not offer seamless throughput when the real beam direction is not identified due to the cell phone detection failure. 
Furthermore, due to the mismatch between the pre-defined discrete beam direction and the real direction, codebook-based beam management techniques fail to provide reliable throughput in many parts of the service area.

\begin{figure}[t]
	\centering
    \includegraphics[width=1.0\columnwidth]{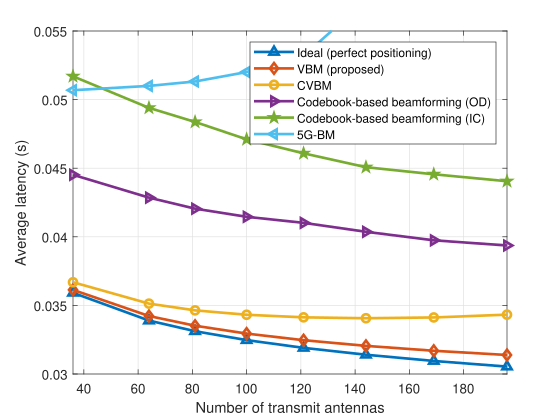}
    \vspace{-0.6cm}
	\caption{Average latency as a function of the number of transmit antennas.}
	\label{fig:latencyresult}
  \vspace{-0.5cm}
\end{figure}

In Fig. 14, we plot the average transmission latency, which is defined as the delay to transmit the data packet~\cite{kjsuh_access}, as a function of the number of transmit antennas.
In our simulations, we set the data packet length $F_{\text{length}}$ to 1 Gbits and then compute the average transmission latency as $D_{\text{avg}} = \frac{F_{\text{length}}}{WR_{\text{avg}}}$.
We also set the number of beams used for sweeping in 5G-BM to $36$.
As shown in Fig.~\ref{fig:latencyresult}, one can see that the transmission delay of VBM is much smaller than those of the conventional techniques.
For example, when 196 transmit antennas are employed,
the latency of CVBM is way higher than that of VBM since the angle mismatch caused by CVBM's cell phone detection failure results in the latency increase.
While it is true that the average latency of VBM decreases steadily with the number of antennas, latency of 5G-BM increases rapidly. 
For instance, when 36 antennas are used, the average latency of VBM is 36$\,$ms, which is 29\% smaller than that of 5G-BM (50.7$\,$ms). 
Whereas, when using 121 transmit antennas, VBM achieves 40\% latency reduction over 5G-BM.
This is mainly due to the beam sharpening effect associated with the growing number of transmit antennas.
Recalling that 5G-BM transmits very sharp discretized beams, one can see that the misalignment between the pre-defined beam direction and the real one will increase the transmission latency considerably.

We mention that there are corner cases where object detection from the captured image might not work properly. 
For example, if the target mobile is visually blocked by obstacles, then the BS cannot identify the beam direction.
To deal with this issue, one can consider several approaches.
One approach is to exploit the hybrid use of VBM and conventional RF transmission.
For instance, one can selectively use VBM in situations where the target mobile is visible and return to the conventional approach (beam index feedback and codebook-based beamforming) when the blockage is detected via VBM.
Another approach is to exploit a multi-modal sensing information obtained from multiple sensor
devices (e.g., radar, ultrasonic sensor, infrared and thermographic cameras). For example, one can simultaneously use radar and camera to detect the target mobile blocked by obstacles.
By exploiting these approaches, one can significantly reduce the user outage probability.

To judge the transfer ability (effectiveness of the proposed technique in different
environments), we evaluate the cell phone detection performance of the VOMTC-trained object detector using a newly collected dataset (see Table III).
We refer to this recently collected dataset as VOMTC-V2 (see https://github.com/islab-github/VOMTC).
From our experiments, we confirm that the localization performance of our object detector in VOMTC-V2 is comparable to that in VOMTC.
Based on the localization results in Table III, we plot the average data rate of VBM as a function of the number of transmit antennas (see Fig. 15(a)).
We observe that the data rate performance of VBM in VOMTC-V2 is also comparable to that in VOMTC.
For example, when using 64 transmit antennas, VBM in VOMTC-V2 achieves about 30.6 bps/Hz, which is very close to the average data rate achieved in VOMTC (31.4 bps/Hz).

To demonstrate the efficacy of the proposed VOMTC-aided wireless communications framework in IRS channel estimation, we have newly investigated its IRS-UE channel estimation performance in terms of NMSE (see Fig. 15(b)).
In our framework, the VOMTC-trained object detector identifies the visible UE and its location in an image captured by the imaging sensors installed on the IRS. 
Using the acquired location of the UE, the geometric channel parameters of the IRS-UE channel can be readily obtained. 
As shown in Fig. 15(b), the channel estimation performance of the proposed framework is similar to that of the oracle-LS scheme, which assumes that the location of the UE is perfectly known. 
For example, when using 112 IRS reflecting elements, our framework achieves an NMSE of about 0.016 dB, which is very close to the NMSE achieved by the oracle-LS scheme (0.015 dB).
\begin{table*}[t]
    \centering
    \caption{Object localization performance evaluated on the VOMTC-V2 dataset.}
   
    \resizebox{2.0\columnwidth}{!}{
    \begin{tabular}{c|ccc|cc}
        \hline
        &
        \multicolumn{3}{c}{Cell phone} & \multicolumn{2}{|c}{Localization error}  \\
           & Precision (\%) & Recall (\%) & F1 score (\%) & Azimuth Angle (rad) & Elevation Angle (rad) \\
         \hline
        VOMTC-trained Object Detector & \textbf{94.93}    & \textbf{76.16}  & \textbf{84.52}   & \textbf{0.0977}  & \textbf{0.0987}        \\
        EfficientDet-D8  & 94.03   & 73.25  & 82.35 & 0.1094 & 0.1102  \\
         \hline
        \end{tabular}}
    \label{tab:vobemv2_result}
\end{table*}
\begin{figure*}[t]
	\centering
    \subfloat[]{\includegraphics[width=1.0\columnwidth, height = 6.5cm]{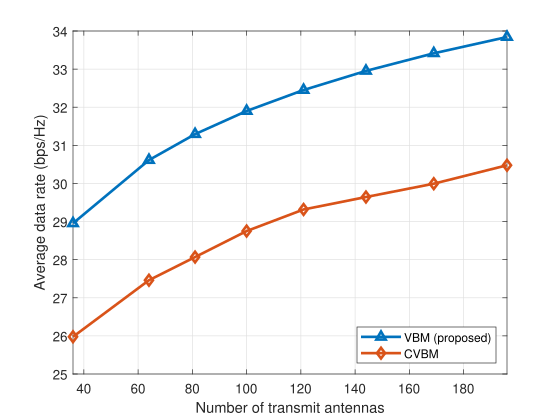}}
     \subfloat[]{\includegraphics[width=1.0\columnwidth, height = 6.5cm]{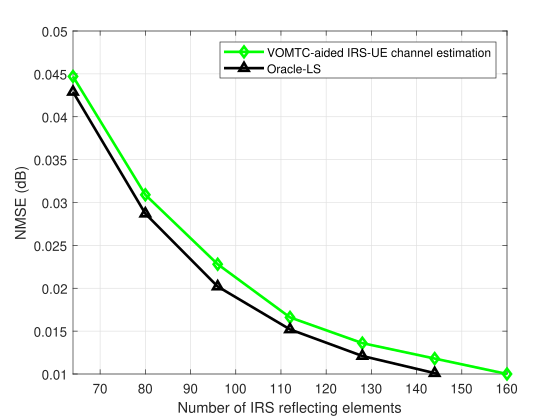}}
       \vspace{-0.2cm}
	\caption{(a) Average data rate as a function of the number of transmit antennas in VOMTC-V2 and (b) NMSE as a function of the number of IRS reflecting elements.}
	\label{fig:newenvironment_IRS}
  \vspace{-0.5cm}
\end{figure*}

\section{Conclusion}
In this paper, we proposed a large-scale vision dataset called VOMTC targeted for the CV-aided wireless applications.
From the numerical experiments on wireless device identification and mmWave/THz beamforming, we observed that the VOMTC-trained object detector outperforms conventional object detectors trained by the commonly used dataset.
It is now clear that AI will play a key role in 6G systems and there will be a variety of wireless applications for VOMTC such as IRS control, blockage detection, and channel/crowdness modeling.
To download the VOMTC dataset, selection code, and the object detection training/evaluation codes discussed in this paper, check out https://github.com/islab-github/VOMTC.

\bibliographystyle{IEEEtran}

\end{document}